\documentclass[11pt]{article}  
\usepackage{menuproc}
\usepackage{cite}
\usepackage{epsfig}
\usepackage{amsmath,amssymb}
\begin{document}
%
%
%

\titlematter{Resurrecting the KH78/80 partial wave analysis}%
{P. Piirola$^a$, E. Pietarinen$^{a,b}$, and M.E. Sainio$^{a,b}$}%
{$^a$Department of Physics,\\
 $^b$Helsinki Institute of Physics,\\
     P.O. Box 64, 00014 University of Helsinki, Finland}%
   {Most of the data from the meson factories were available only
     after the $\pi N$ partial wave analysis of Koch and Pietarinen
     \cite{kochpie} was published over 20 years ago.  Since then, both
     the experimental precision and the theoretical framework have
     evolved a lot as well as the computing technology.  Both the new
     and the earlier data are to be analysed by a highly modernised
     version of the earlier approach.  Especially the propagation of
     the measurement errors in the analysis will be considered in
     detail, visualisation tools will be developed using the
     Python/Tkinter combination \cite{python}, and the huge data base
     of experiments will be handled by MySQL \cite{mysql}.}
%
%

\section{Introduction}

About 20 years ago Koch and Pietarinen performed an energy-independent
partial-wave analysis on pion-nucleon elastic and charge-exchange
differential cross sections and elastic polarisations for laboratory
momenta below 500 MeV/c incorporating the constraints from fixed-$t$
dispersion relations as well as crossing and unitarity (the KH78 and
KH80 analyses) \cite{kochpie}.  Since then, however, new low-energy
data have emerged in all charge channels: examples of recent high
precision results for the differential cross sections are given in
refs. \cite{pavan,frlez,janousch}, polarisation parameter in refs.
\cite{gaulard,hofman,wieser} and the spin rotation parameter in ref.
\cite{supek}.  There are also some new measurements of integrated
cross sections \cite{kriss}.  Especially the high precision
measurements of the hadronic level shift and width on pionic hydrogen
and deuterium \cite{PSI}, giving information of the pion-nucleon
interaction just at the threshold, have opened a completely new
chapter in the study of the low-energy pion-nucleon interaction.
Another direction where significant advances have taken place is the
theoretical framework where we study the low-energy pion-nucleon
interaction.  The tool, chiral perturbation theory ($\chi$PT), has
been developed in the 80's and 90's and the work continues.  The
development of $\chi$PT motivates a new partial wave analysis from the
theoretical point of view --- on one hand strong interaction physics
is becoming a precision science also in the low-energy region, on the
other hand, there is need for $\pi N$ phenomenology to fix some of the
low-energy constants appearing in the meson-baryon lagrangian.  See,
for example, the talk of Mei\ss ner in these proceedings
\cite{meissner}.

It is our goal to make use of the new data in an analysis which, in
addition to the requirements of analyticity, crossing and unitarity,
includes the constraints from chiral symmetry.

\section{The old KH78/80 analysis}

The aim of the KH analysis was to determine the amplitudes satisfying
several conditions:
\begin{enumerate}
  
\item The amplitudes had to reproduce all the data which were available
  at that time, which means $\frac{d\sigma}{d\Omega}$,
  $\sigma_{\mathrm{tot}}$ and $P$ within their associated errors.
\item The solution had to fulfil the isospin invariance.
\item All partial waves had to satisfy the unitarity condition.
\item The crossing symmetry was implicitly assumed, because 
  Mandelstam variables were used.
\item The invariant amplitudes were going to have the correct
  analyticity properties in $s$ at fixed-$t$.
\item The amplitudes at fixed-$s$ were to be analytic in $\cos \theta$ in
  the small Lehman ellipse.
\end{enumerate}
The experimental data is not enough to fix a unique partial wave
solution, but further theoretical constraints are needed.  The
constraints from fixed-$t$ analyticity and from the isospin invariance
are strong enough to resolve the ambiguities \cite{hohler}.

\section{Three stages of the analysis}
\subsection{Fixed-$t$ analysis}

The old KH analyses consist of three phases: fixed-$t$ analysis,
fixed-$\theta$ analysis and fixed-$s$ analysis, which were iterated
until the results agreed up to about 3 \%.  The fixed-$t$ analysis was
carried out at 40 $t$-values in the range from zero to $-1.0$ GeV$^2$.
The analysis would be too complicated, if one were working with
dispersion integrals, so the expansion \cite{pietexp1,pietexp2}
\begin{equation}
  \label{pietexp}
  C(\nu, t) = C_N(\nu,t)+H(Z,t) \sum_{i=0}^n c_i Z^i
\end{equation}
was used for $t$-values smaller than $-4m\mu$, ($\nu = (s-u)/4 m_N$).
In the expansion the nucleon pole term $C_N(\nu,t)$ is treated
separately, and the sum is multiplied by a factor $H(Z,t)$, which
describes the expected asymptotic behaviour.  The essence of the
expansion is that the sum is written in terms of functions~$Z$, which
have the correct analytic behaviour, i.e. it is {\em not} a polynomial
approximation, but a series presentation of an analytic function,
which is just truncated at some reasonable point (ca. $n=50$ or
$n=100$), because infinite accuracy is impossible.  The condition of
smoothness and the compatibility with the data constrain the terms
with large index $i$ to be negligible \cite{pietexp1,pietexp2}.

The expansion coefficients $c_i$ are determined by minimising
\begin{equation}
  \chi^2 = \chi^2_{\mathrm{data}} + \chi^2_{\mathrm{pw}} + 
  \chi^2_{\mathrm{penalty}} \;.
\end{equation}
Here $\chi^2_\mathrm{data}$ comes from the experimental errors,
$\chi^2_\mathrm{pw}$ belongs to the deviation from the fixed-$s$
partial wave solution and the last term is used to suppress large
values of the higher coefficients of the expansion.  In practice, the
analyticity constraints cannot be used without smoothing the data.
The aim is, of course, to smooth out the statistical fluctuations
without distorting the physically relevant structures.

\subsection{Fixed-$\theta$ analysis}

The fixed-$t$ constraint is often used only for $t$ values from zero
to about $-0.5$ GeV$^2$, because the partial wave expansions for the
imaginary parts of the invariant amplitudes do not converge for large
$|t|$.  In the range $t \in (-0.5, -1.5]$ GeV$^2$ the truncated
partial wave expansions can still be reasonable approximations, but
for $t$ values below ca. $-1$ or $-1.5$ GeV$^2$ the fixed-$t$
analyticity cannot be applied anymore.  So another analyticity
constraint is used to cover the rest of the angular range at
intermediate and at high energies.  The calculation was made at 18
angles between $\cos \theta = -0.9 \ldots 0.8$.  Analysing methods are
the same as in the fixed-$t$ analysis: the expansion method is used
and the coefficients are fixed by minimising $\chi^2$, i.e. by fitting
to the data and to the fixed-$t$ solution.

\subsection{Fixed-$s$ analysis}

The third stage, the fixed-$s$ analysis, is a standard phase shift
analysis in the sense that the partial waves are fitted to the data.
On the other hand, it is not the usual one, because the partial
waves are fitted also to the fixed-$t$ and to the fixed-$\theta$
amplitudes.  Now 92 momentum values were selected from the energy
range $0\ldots200$ GeV/c, 6 of the momenta were above 6 GeV/c.  Again,
the coefficients were fixed by minimising $\chi^2$ which now included
also a term suitable to enforce unitarity.

\section{Treatment of the data}

The electromagnetic corrections proposed by Tromborg et al. at momentum
values below $0.65$~GeV/c \cite{tromborg}, were applied to the data.  At
higher momentum, only the one-photon exchange correction was applied,
taking into account the Coulomb phase.  In all three different analyses,
the data were shifted to the selected energy bins (i.e. the selected
values of $s$, $t$ or $\theta$) using the previous solution of the
iteration to calculate the correction.  Some data points requiring too
large a momentum shift were omitted.  The normalisation of some data sets
had to be corrected to guarantee a smooth extrapolation to the forward
direction, and to the input for the forward amplitude.

\section{Life after KH80}

The latest KH phase shift analysis was finished in 1980.  After that
there has been very accurate measurements of pionic hydrogen level
shift and width, $\frac{d \sigma}{d \Omega}$ has been measured with
good accuracy, many spin rotation parameter measurements has been done
as well as polarisation parameter measurements.  Also, some integrated
cross section measurements have been performed.  The newer data has
never been analysed by the methods of Koch and Pietarinen, and for
example the results of Pavan et al. \cite{pavan} are not compatible
with the results of the old analyses.  So, an updated version of the
analysis is certainly needed.

\section{The code of Pietarinen}

The original code was made for the NDP Fortran compiler, which runs
under MS-DOS.  The code needs to be ported to UNIX.  We have tested
the code in an old MS-DOS machine, and most of the main tasks seem to
be working correctly.  What is still needed, is a modification of the
$s$-plane conformal mapping.  This has effects on all routines, which
are related to the fixed-$t$ analysis.  Also some modifications are
needed to be able to study the isospin analysis.

The code base is divided into several parts:
\begin{itemize}
\item There is a program for comparing the partial wave solution to
  existing data.  It simply plots the data and the solution in the
  same picture, and allows the comparison of different data sets to
  the solution.
\item The second part is for shifting the experimental data points
  into the fixed-$t$ bins.  The earlier solution is used for
  interpolation, and those data points which are to be shifted too
  much are rejected.
\item The next part is for making a starting value for the fixed-$t$ expansion.
\item One program is for the iteration to adjust the fixed-$t$ amplitude
  to the experimental data and to the current solution.
\item The main part of the program makes the actual partial wave
  analysis and adjusts the solution simultaneously to the data and to
  the fixed-$t$ amplitudes.
\end{itemize}

\section{The current status}

\subsection{Porting the code}

During the porting process the most extensive work is needed for
writing the graphical user interface, the data base engine and the
plotting routines.  The graphics routines of the original code were
impossible to get working under UNIX, so we decided to use the
Python/Tkinter combination for the GUI, and the Python/Gnuplot
combination for plotting routines.  The old code was reused as much as
possible, but many parts still needed almost complete rewriting.  We
decided to write all the new code in Fortran 95, so, at the moment,
most of the calculation engine is written in Fortran 77 and some parts
in Fortran 95.  All routines of the old code were modified to take
almost all input from the stdin and to write output to stdout, so it
should now be possible to change the whole GUI with a reasonable
amount of work, whenever it becomes necessary.

\begin{figure}[t]
\centerline{\epsfig{file=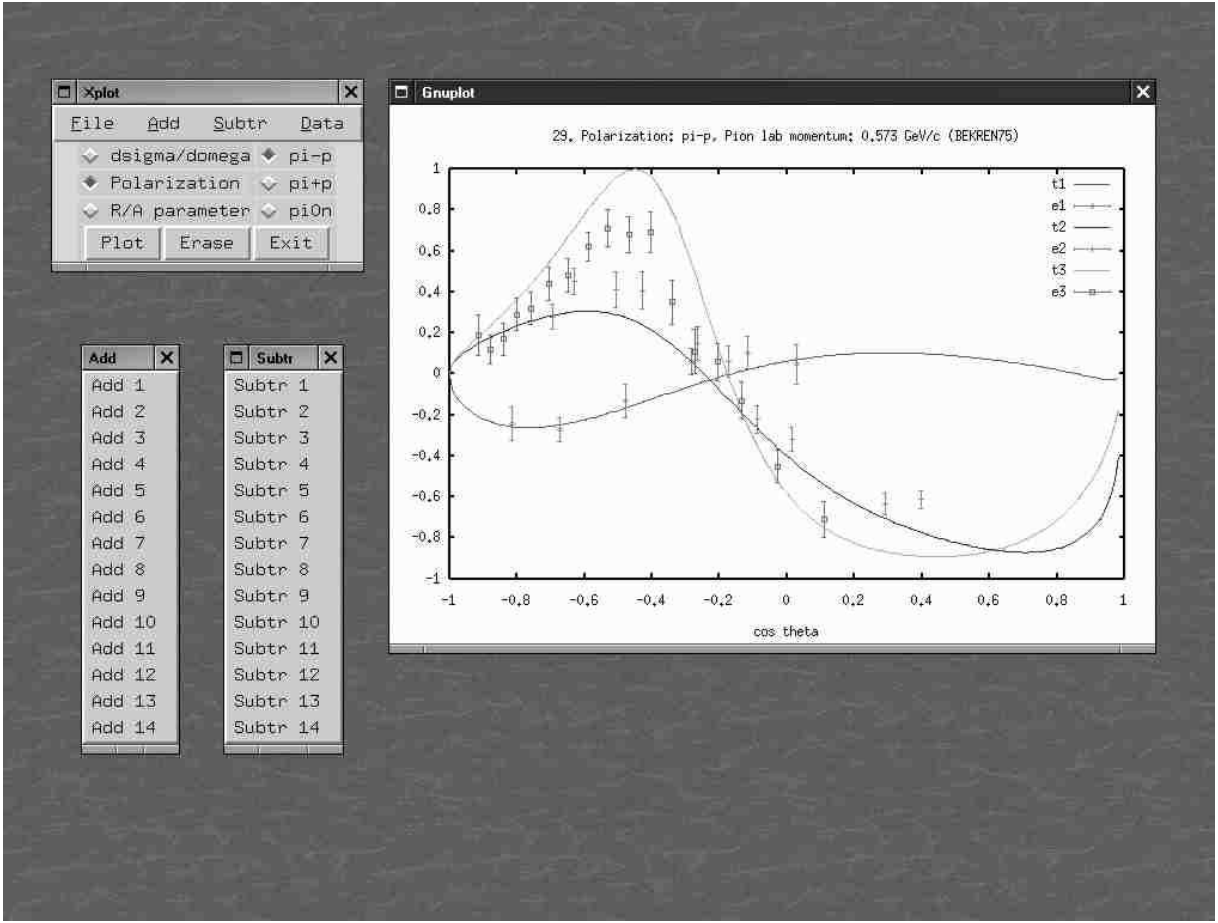,width=.976\textwidth, silent=,clip=}}
\caption{\label{xplot}
  Screen shot of the plotting program.  In the plot there are three
  different measurements at different momentum values ($0.192$~GeV/c,
  $0.425$~GeV/c and $0.573$~GeV/c\,) \cite{alder83,alder78,bekren75},
  and the corresponding partial wave solutions --- so it is possible
  to compare the goodness of the different data sets by the ``eye
  ball'' method.  The scattered particles and the reaction can be
  chosen by the radiobutton widgets.  The different data sets are
  selected by the same method as in the original code of Pietarinen,
  i.e. by moving up or down in the data base with {\tt Add} or {\tt
    Sub} buttons.  The method is quite crude, and it is going to
  change when the data base is converted to MySQL format.}
\end{figure}

\subsection{The current status}

At present, the plotting program, the interpolation program and the
program calculating the starting value of the fixed-$t$ analysis are
ported to UNIX, and all the functionality of the original versions is
implemented (fig. \ref{xplot}).  The part, which iterates to adjust
the fixed-$t$ amplitudes, the partial wave solution, and the
experimental data, compiles OK but the GUI is still under
construction.  The heart of the whole program, the part making the
actual analysis, is still in a completely untested stage.

\subsection{Comparison of the numerics}

We have compiled the code with different compilers running on
different operating systems in order to check the stability of the
mathematical subroutines\footnote{%
  We have tried the old NDP Fortran compiler for MS-DOS, the Compaq
  Fortran in Digital UNIX, the Lahey/Fujitsu Fortran for Linux, and
  different versions of the GNU/Linux Fortran.}. Comparing the results
of the routines compiled by different compilers showed that the
routines are {\em not} currently stable enough for production use.

For illustration, the interpolator part of the program calculates
173670 interpolated data points.  When comparing the results of
routines compiled by GNU Fortran to those calculated by NDP Fortran,
one notices that 96\% of the new data points agree up to 0.01\%, but
in some cases there are significant discrepancies: namely 73 data
points of the 173670 differ by more than 1\%, and in the worst case the
difference is 28\%.  The cause of these discrepancies is unknown when
writing this.

\section{The next phase}

After chasing the bugs and finding the reasons for the numerical
unstabilities, we are hoping to find a better way to handle the
propagation of the experimental errors than that used in the old KH
analysis.

Because of the size of the data base, and because of the discrepancies
in the different data sets, the visualisation of the data and the
partial wave solutions is essential.  Also, during the analysis, the
program is used a lot, so the graphical user interface has to be easy
and efficient to use.  For making the choice of the data sets as
easy as possible, we have plans to convert our text file data bases to
MySQL format.

We intend to get the first preliminary results by the end of the year.



\acknowledgments{We wish to thank A.M. Green for useful comments on the
  manuscript.  We gratefully acknowledge financial support by the
  Academy of Finland grant 47678, the TMR EC-contract CT980169, and
  the Magnus Ehrnrooth foundation.}


\end{document}